# Expanding Conservation Science through Emerging Interdisciplinary STEM Fields

Andrew Schulz[1,+,*], Adam S. Gouge[2+], Christine L. Madliger[2s,+,*]

[1]Mechanical Engineering Department, Georgia Institute of Technology, Atlanta, GA USA 30318
[2]Department of Biology, Algoma University, Sault Ste. Marie, Ontario, Canada P6A 2G4

**Keywords**: Zoological Research, Conservation Biomechanics, Conservation Physiology, Conservation Technology, Interdisciplinary Collaborations

+Designates equal amounts of work
Corresponding Author(s):

Andrew Schulz
akschulz@gatech.edu

Christine Madliger
christine.madliger@algomau.ca

**Abstract:**

Conservation science is an interdisciplinary field that primarily draws on knowledge from the natural sciences, social sciences, and humanities to inform policy, planning, and practice. Since its formalization as a discipline, conservation science has also increasingly incorporated tools from integrative biological fields, such as animal behavior, genetics, and, more recently, physiology. Given that the biodiversity crisis constitutes one of the greatest challenges of the 21st century, with tremendous consequences for global sustainability and human health, creating a diverse conservation toolbox is important for addressing complex conservation threats. To assess the integration of three emerging integrative biological disciplines (physiology, biomechanics, and technology) into recent conservation science research, we queried publications from five broad-scope conservation-focused journals from 2010-2022. We found that the proportion of published articles incorporating these integrative biological techniques was low, ranging from 0-4% per year. With only 2.1% of total articles accessing tools or techniques from conservation physiology, conservation technology, and conservation biomechanics, we propose that there is still a substantial opportunity for further integration. We provide a case study for each integrative field to illustrate the capacity for its tools to contribute to positive conservation outcomes. We further outline how each field promotes novel or reimagined opportunities for collaborations. Finally, we discuss the interconnectedness of the three fields and how they can support the continuing expansion of conservation science as an evidence-based, action-oriented discipline through the application of a Challenge-Mechanism-Partnership framework.

**The state of the art of conservation science**

As a result of human activity, large-scale ecological and environmental changes have led to a tremendous loss of biodiversity, with current extinction rates exponentially greater than background rates[1,2]. Biodiversity has seen a 70% decrease in the last 50 years, and significant species declines have been quantified in vertebrates[3], invertebrates[4], plants[5], and fungi[6]. Due to ecological lag, future extinctions are likely to be greater than prior estimates, and current conservation action plans may not address this exacerbated loss of species, suggesting that opportunities to mitigate or substantially decrease future species loss are even more pressing[1]. Given the need for evidence-based conservation action, stocking the conservation toolbox with a wide range of well-validated techniques and approaches is paramount.

As an applied academic discipline, conservation biology has deep roots in ecology, population biology, biogeography, and the management of natural resources[7]. During its formalization, it centered on combining theory with practical experience from forestry, fisheries, wildlife, and parks management[7]. Over time, conservation biology expanded to draw in a broader range of integrative or mechanistic biological disciplines. For example, genetics is a now cornerstone of conservation science[8] and studying behavior is instrumental for many conservation monitoring and captive breeding programs[9]. More recently, the field of conservation physiology was codified[10,11], but has not yet become a common component of conservation-focused textbooks, conferences, or undergraduate courses. Beyond the continued calls for the expansion and refinement of these conservation science sub-fields (e.g., the expansion of conservation genetics to consider epigenetics and genomics[8]; the inclusion of sensory ecology[12] and microbiomes[13] in conservation physiology), there have also been recent suggestions for new integrative and interdisciplinary approaches related to paleobiology[14], synthetic biology[15], technology[9,16], and biomechanics[17], among others. Here, we present a discussion of the value and interconnectedness of three emerging integrative subfields of conservation science: conservation physiology, technology, and biomechanics. We first illustrate the lack of publications in significant conservation journals focused on these topics and then provide case studies to illustrate their potential for tangible contributions. We also present how these emerging fields promote novel or reimagined collaborations outside of the traditional academic conservation-science realm.

**Assessing the state of integration of novel conservation disciplines**

To locate conservation science publications that mentioned terms related to integrative disciplines (physiology, biomechanics, and technology), we completed three separate searches using the advanced search feature on PubMed similar to the method employed by Finni et al[18]. We queried the titles, keywords, and abstracts of journal articles published from 2010 to 2022 in five high-ranking conservation journals: *Conservation Biology*, *Conservation Letters*, *Biological Conservation*, *Animal Conservation*, and *Biodiversity and Conservation*. Our search strings can be viewed in the Supplementary Materials (). We calculated the proportion of articles each year corresponding to each integrative sub-disciplines compared to the total number of articles published across all five journals.

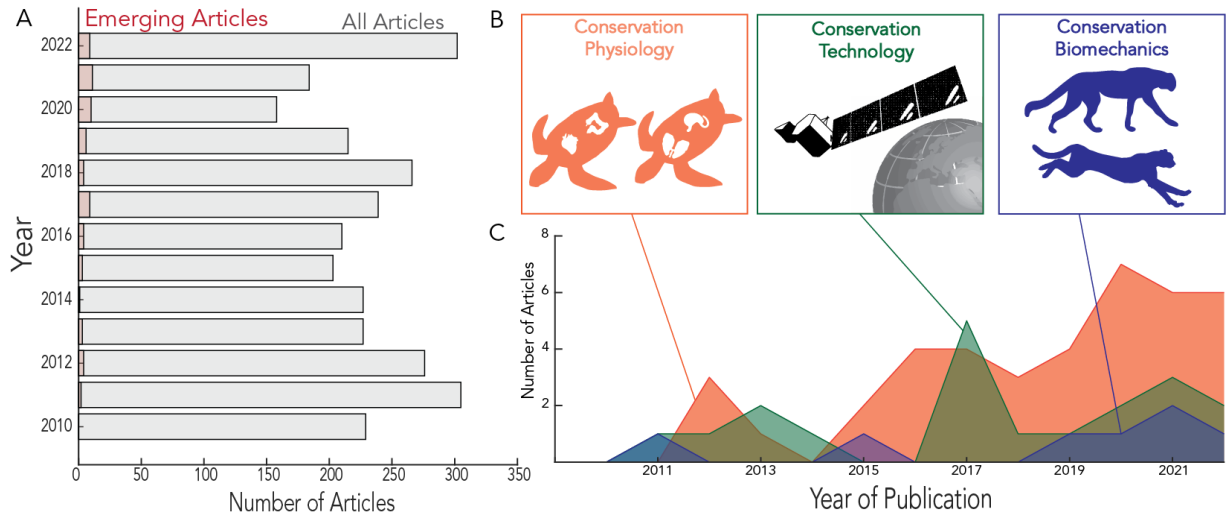

**Figure 1.** (A) Number of articles focusing on technological, physiological, and biomechanical approaches published in five representative conservation science journals (*Conservation Biology*; *Conservation Letters*; *Biological Conservation*; *Animal Conservation*; *Biodiversity and Conservation*) from 2010-2022. (B) Case Studies we are targeting as emerging conservation fields. C) Number of articles focusing on integrative fields (physiology, biomechanics, and technology) in comparison to total articles published.

Of the three emerging integrative fields, articles referencing technology were most common (0-12 publications/year), followed by physiology (0-7 publications/year), and then biomechanics (0-4 publications/year) (Figure 1). Compared to the full complement of articles published in conservation journals, the integrative fields accounted for a small percentage in combination, ranging from 0-4% per year and 2.1% overall (Figure 1). The field of conservation physiology has a discipline-specific journal (*Conservation Physiology*, Oxford), which could partially account for the low proportion of articles occurring in non-discipline-specific conservation journals. However, given that the journals we queried are high-impact venues within the broader field of conservation science, including a number of society journals, they ideally should reflect the full range of tools and approaches available.

**Three avenues for collaborative, action-oriented integration**

Conservation science is inherently collaborative; the first International Conference on Conservation Biology occurred in 1978 and brought together zoo managers with academics and wildlife conservationists[7]. Given the state of the art in conservation science, these traditional collaborations that naturally stem from the ecology, behavior, and microbiological fields have been well-maintained. Many of these collaborations are operating in the field, and therefore, collaborations with conservation organizations, wildlife managers, field surveyors, species trackers, and community scientists are common in the terrestrial environment. Within aquatic conservation science, fishmongers, fishery workers, and coast patrollers can also help with species capture to collect DNA samples. These traditional collaborations are proven to help projects create a more significant and holistic impact, but for many of the more novel conservation sub-fields discussed in this paper, we must expand these collaborations. Here, we present a case study for each of three emerging biological sub-disciplines of conservation science: conservation physiology, technology, and biomechanics. For each, we discuss important collaborations that are essential to their success.

Many of the techniques discussed in the following sections utilize newer collaborations, including the nearly 1.1 billion objects held in natural history museums throughout the globe. In other cases, the success of the endeavor relied on professional connections between not-for-profit centers that rely on volunteer effort, veterinarians, and academic partnerships. Additionally, researchers are collaborating with zoological organizations to leverage partnerships in novel ways, including bio-inspired design[19], conservation technology[20], or eDNA collection devices[21]. These new and research-expansive zoo-academia research collaborations can be used to inform conservation practices through a potential understanding of welfare, biomechanics, reintroductions, and more.

**Case Study 1 - Physiological assessments support rehabilitation programs for critically endangered sea turtles**

*What is conservation physiology?* Conservation physiology is an action-oriented discipline that employs physiological tools, techniques, and knowledge to characterize diversity, quantify responses to environmental change at multiple biological scales, and inform conservation decisions[10,11]. Conservation physiologists use approaches from various physiological sub-disciplines, such as endocrinology, metabolism and energetics, thermal biology, and immune function, among many others (see toolbox by Madliger et al. 2018[22]). Detailed descriptions of the field have been provided[10,11,23,24] and the tools available can address conservation challenges across taxa, including in organisms as small as bees[25]. As the field has developed, an increasing number of minimally and non-invasive techniques have been validated, and increasing options for obtaining physiological information from wild organisms abound. By contributing to endeavors such as animal welfare and captive breeding, identification and monitoring of threats, managing human-wildlife conflict, habitat restoration, predicting the spread of invasive species, and predicting tolerances under climate change scenarios, conservation physiology can support broader scientific aims[24]. Specifically, conservation physiology can address some of the goals associated with the United Nations Sustainable Development Goals and the Scientists' Warning to Humanity's calls to action (see summaries in Cooke et al., 2020[23] and Madliger et al., 2021[26], respectively).

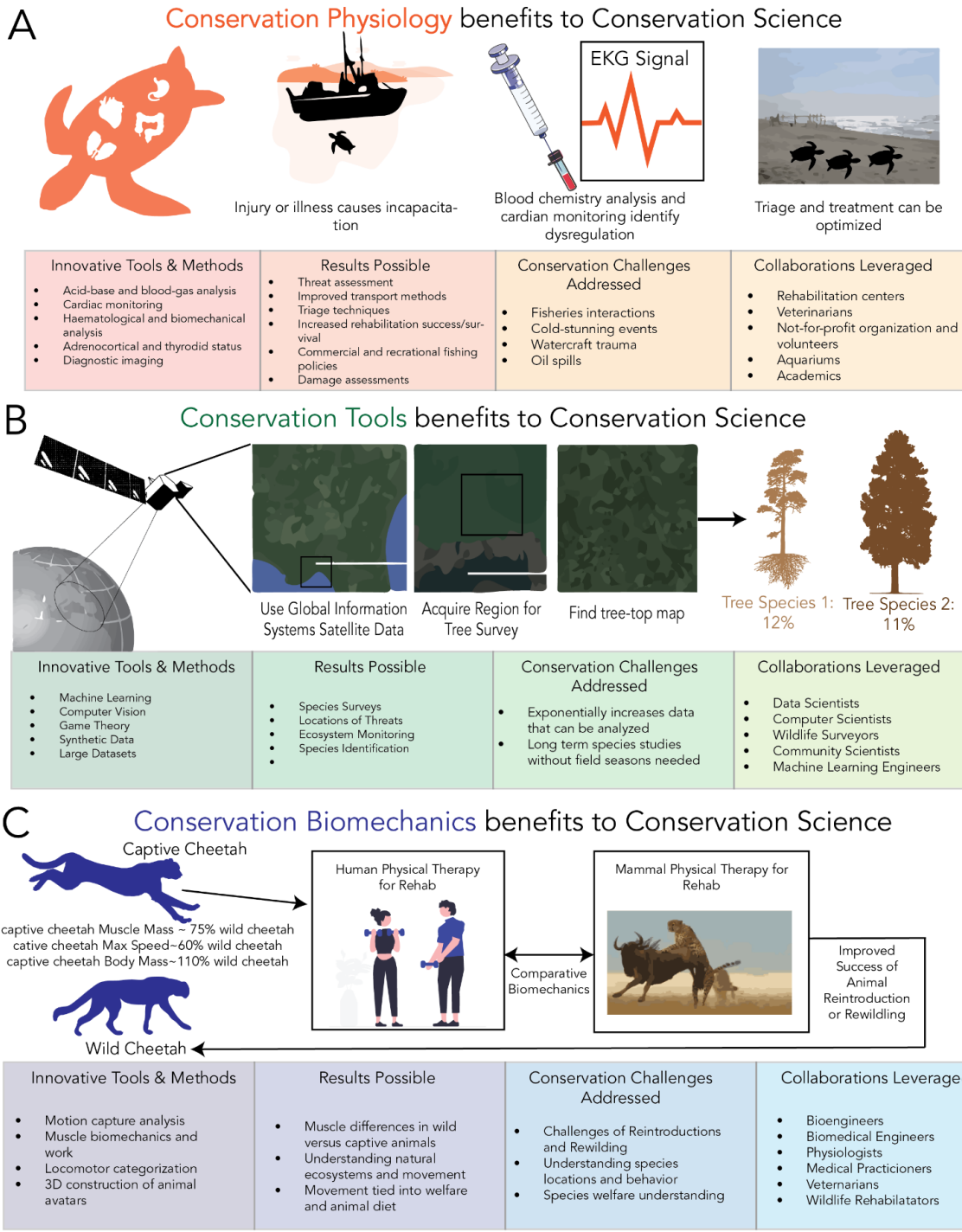

**Figure 2.** Overview of the physiological techniques that can be used to address conservation challenges in critically endangered sea turtles. The realized conservation outcomes from the incorporation of physiological information are also highlighted, along with the types of collaborators that allow the collection, translation, and incorporation of physiological knowledge into conservation action. A) Summarized from a larger discussion of the intersection between conservation physiology and sea turtle rehabilitation in Innis & Dodge, 2020[27]. B) Summary of a larger paper on AutoArborist Dataset by Beery et al[28] detailing detection of tree species in different areas from overhead images. C) Summary of an emerging conservation field of conservation biomechanics highlighting the cheetah and a rehabilitation pipeline.

*What can conservation physiology accomplish?* The potential for conservation physiology to directly inform conservation-relevant actions is well-illustrated by physiology-informed rehabilitation efforts for endangered sea turtles (see full review of topic in Innis & Dodge, 2020[27]) (Figure 2A). Globally, sea turtle populations have declined due to coastal development, hunting, boating, pollution, fisheries interactions, and the direct harvesting of eggs[29]. Sea turtles that become ill and/or injured due to cold-stunning events, oil spills, interaction with boats, or incidental fisheries capture (i.e., bycatch) can be treated at rehabilitation facilities[27]. When large numbers of cold-stunned turtles require treatment simultaneously, triage decisions must sometimes be made due to limited resources[27]. Through the consideration of physiological data, the New England Aquarium (Massachusetts, USA) has been able to apply a mortality prediction index for Kemp's ridley turtles (*Lepidochelys kempii*) to direct resources and effort toward the individuals with the best prognoses[30]. Specifically, the index uses measurements of blood pH, oxygen content, and potassium levels to assess physiological dysfunction and provide a mortality prediction[30]. The application of this type of index was expanded to also evaluate turtles impacted by the 2010 Deepwater Horizon oil spill, improving triage, directing treatment, and benefitting overall rehabilitation outcomes to result in an over-90% survival rate of treated oiled turtles[31].

In some locations, turtles must also be transported large distances to or from rehabilitation facilities prior to release, leading to physiological stress associated with handling, confinement, physical motion, and temperature variation[32]. Monitoring of corticosterone, glucose, potassium, and other blood parameters associated with the stress response indicated that allowing transported Kemp's ridley sea turtles to recover for 6 hours in saltwater tanks at their release site can allow for physiological recovery[32]. Beyond improving protocols for rehabilitation, the growing body of physiological work in sea turtles has informed policy and planning. The data collected during the Deepwater Horizon spill contributed to estimating overall sea turtle losses, which was then used for the natural resource damage assessment[33] as well as in the legal proceedings that led to a $20 billion USD settlement, a portion of which was directed to sea turtle conservation projects[34]. Further, the associations of blood biochemistry parameters (obtained from point-of-care devices) with sea turtle health status has contributed to policy decisions regarding soak time limits for coastal gillnets aimed at minimizing bycatch[35].

*Who is involved?* The collection and application of physiological data for the benefit of rehabilitation and recovery of sea turtles was possible because of multi-partner collaborations between veterinarians, sea turtle biologists, volunteers and other personnel at rehabilitation centers, and aquariums. Veterinarians routinely work with panels of physiological measures that can be integrated with conservation goals to monitor, manage, and plan recovery strategies for at-risk wildlife. In particular, this case study illustrates the positive outcomes that can be generated when biologists work directly with conservation-focused organizations to problem-solve (e.g., to improve efficiency and resource allocation, create more effective protocols for rehabilitation, and inform assessments).

**Case Study 2 – Conservation technology allows for larger data sets to be compiled in shorter times with global collaborations through open science**

*What is Conservation Technology?* Hardware and software developments and the field of conservation technology, or recently also phrased conservation tool creation, are beginning to have a greater impact on conservation science[16]. The primary objective in this field is to develop purpose-driven technology and software instead of co-opting the historical opportunities of hardware and software[36]. Items like camera traps are not designed for ecologists, but ecology took the opportunity to apply this tool. Two of the most significant impacts in the emerging collaborations of hardware and software development with conservation science are open-source science and machine learning tools.

The Free-and open-source Hardware, or FOSH, seeks to address the inaccessibility and expense of conservation science items. FOSH is just one example within the overall open science movement which recently has been reviewed by Bertram et al[37]. Examples of open-source hardware includes field kits, DNA testing stations, camera traps, wildlife trackers, and more. Many scientists are working on developing items that have low cost, are easy to use, can be modular for different field sites, and are purpose-built for specific conservation science scenarios. Machine learning (ML) is the concept of recognizing and analyzing patterns in data[38]. A supervised ML algorithm can take in large quantities of inputs, such as images of different species and their expected results, in the form of labels of the species names. In many conservation science fields, especially wildlife and population monitoring, there can be thousands of hours spent on labeling data from GIS or camera trap images[39]. With ML, models can be trained to accomplish these tasks in hours instead of months[40]. Additional examples including MammalNet[41], ModelZoo[42], Wildbook[43], and iNaturalist[44] are leveraging the work of community members and community-backed science to address these challenges.

*What can conservation technology accomplish?* A prime example of how conservation technology is contributing to more than just animal-focused conservation science is the AutoArborist project, which uses advanced computation techniques for urban forest monitoring (Figure 2B)[28]. The AutoArborist project works to study computer vision challenges as well as address conservation and environmental challenges using the study scenario of urban forests. Forests are inherently complex to monitor as they can span millions of acres across multiple countries, and massive data information to measure the change in the ecosphere over time is often required[45]. Traditionally, monitoring is focused on species surveys, including nature walks and documenting and counting trees and items like urban encroachment on an area. In Los Angeles (US), a recent census costs $2 million and took 18 months[46]. Computer vision and machine learning techniques utilized a training set to generate a model, and then AutoArborist created a *tree census* pipeline consisting of 2.6 million trees spanning 23 cities across the United States[28]. Specifically, the biodiversity of plant life in a country-scape of the US has incredible biodiversity, with the data set in the US having 344 unique Genera.

The generation of this type of data set has allowed other researchers to apply it for different types of species monitoring, ecosystem health assessments, and habitat modeling across the entire US (Figure 2B). Computer vision and machine learning techniques like AutoArborist also exist in other realms of the computer vision space, including items like human-body pose detection[47], 3D animal model reconstruction[48], individual species identification[49], among others.

*Who is involved?* Using open-sourced data sets and machine learning as conservation tools helps increase the amount of data processing possible in conservation science. Utilizing these tools in creative ways allows research to progress and extend beyond that of the traditional fields of ecology and behavior. This dataset is open and accessible to ecologists, computer scientists, engineers, and all who would like to utilize it, and programs are becoming more common that are working on expanding the fields of computer vision for conservation to graduate ecologists[50], engineers[51,52], and undergraduate biologists[53].

**Case Study 3 – Proposing biomechanics as an emerging field to assist in more successful reintroductions**

*What is conservation biomechanics?* In this paper, we define conservation biomechanics as applying the field of comparative biomechanics and animal movement to inform wildlife conservation practices such as enrichment, reintroductions, and more. Currently, there is little to no use of the biomechanics field in conservation science, as shown by published literature in conservation science journals (Figure 1). In fields like human physiology, benchmarks exist using physiology and movement for rehabilitation[54], and we propose that similar benchmarks and ideas could be utilized in the conservation science sector for biomechanics. Of the fields discussed in this perspective this is the least common as the connection between biomechanics and behavior is new when we look at non-humans.

*How is conservation biomechanics emerging and what can it accomplish?* There are many conservation challenges throughout the globe that vary based on the species of concern, but an overarching challenge across the conservation space is adequate data. These data are even more sparse on species that are hard to locate, track, or investigate. Reintroduction, in particular, is an immensely complex science. In scenarios where captive animals are released, individuals must be vetted for their ability to fight disease, find a mate, and successfully forage. For predators, this is incredibly challenging, as food is not as simple as foraging for fruits on the tree, but instead involves hunting other moving targets[55]. Currently evaluation criteria for reintroduction of species is often focused on locating a suitable environment where that species is in decline and needs to be increased[56], but overall reintroductions success, especially in isolated populations (such as many large carnivores), can be difficult to achieve and dependent on several regional factors[57]. Therefore, we provide perspectives of an additional aim in successful reintroductions: the use of individual animals' biomechanical benchmarks.

Often baselines biomechanical benchmarks are unknown. When viewing species such as the cheetah, there are vast differences in morphology between captive and wild individuals. For example, the less active lifestyle of captive cheetahs has been shown to alter skeletal muscle fiber size distribution resulting in a pattern that is consistent with general myopathy (Figure 2C)[58]. Further, the overall proportion of body mass accounted for by the locomotor (limb and back) muscle is higher in wild cheetahs compared to their captive counterparts[59]. These morphological differences, along with potentially different physiological constraints in captive individuals due to discrepancies in the nutritional quality of their diets[60], likely contribute to the generally lower top speeds recorded in captive compared to wild cheetahs[61]. The overall lower mobility and the lack of use of behaviors to subdue prey in captive animals also influences the morphology of the forelimb, the development of which is integral to being able to unbalance their prey while traveling at high speeds[62].

Humans that succumb to injury have specific benchmarks for their biomechanical ability that are evaluated by a physical therapist. These benchmarks include physiological measurements (muscle and bone health) and biomechanical benchmarks (running, walking, or swimming). To craft these benchmarks, biomechanists perform experiments to measure physiological output, biomechanical health, and skeletal health. These similar types of benchmarks traditionally do not exist for captive animals, but we have both sides of the spectrum for some species. Cheetahs in captivity have difference of physiological health and skeletal health with difference of muscle mass compared to that of wild cheetahs[59]. Additionally, the biomechanical benchmarks for sprinting cycles are lower compared to that of wild cheetahs[61]. Some of these benchmarks exist in the domesticated dog literature with healthy balances of physiology[63] and biomechanical output[64] in the scope of veterinarians, but currently has yet to reach non-domesticated or agriculture-housed species.

Cheetahs represent an ideal case study as there are known behavioral between captive and wild individuals. We aim to more broadly introduce authors to considering conservation biomechanics as an additional facet of reintroduction science. In addition to examining more traditional metrics like ethograms and/or physiological indicators of health and welfare prior to re-wildling, reintroduction teams can also look at biomechanics of movement in captive individuals and compare them to their wild counterparts to help create a benchmark for further health evaluation. We believe that incorporating biomechanics into the conservation framework of reintroductions, as well as other conservation efforts such as captive animal welfare and rehabilitation, will lead to more longitudinal successes. A challenge of this would be wild data collection, but as described in Case Study 2 herein (conservation technology) there are large data sets of several species that could be leveraged to foster effective conservation biomechanics benchmarks.

*Who is involved?* Overall, this field benefits from work with Zoological and Aquaria organizations or sanctuaries that are working on reintroduction and rehabilitation. There are several reintroductions of characteristic species, and creating benchmarks through zoological enrichment could allow for reintroductions that are more successful by testing predators' prey-capture mechanisms and prey escape maneuverability. Collaborations with Zoological organizations for academic units provide many benefits on both sides[19]. Additionally, one of the novel parts of conservation biomechanics is that it lends new organisms for study. We often think of model organisms in the biological sciences, but it is important for researchers to look at a wide variety of organisms, given that they might have specializations that allow for interesting and progressive scientific innovations[65].

**A framework to identify emerging approaches that benefit conservation science**

Science is constantly in flux with the advancement of new techniques, technologies, discoveries, and challenges. Therefore, to conclude this perspective, we aim to encourage researchers to think about identifying new collaborations with non-traditional conservation science fields and we urge researchers from non-traditional fields to seek out ways to apply their research to conservation science. In doing so, there is the potential to advance biodiversity understanding and push forward novel tools in the broader field of conservation science. While we have introduced three distinct case studies, it is important to remember that these fields are not entirely independent of one other. For example, in the conservation biomechanics case study, the cheetah accelerations and velocities reported were measured using IMU sensors with online data repositories and muscle physiology arguments, and therefore represent a combination of physiology, technology, and biomechanics techniques. There are also opportunities for work within the realm of conservation physiology to continue to be expanded by technologies, such as implanted telemetry tags and miniaturized physiological sensors. Beyond the connections we have highlighted, it is important that we continue looking for potentially conservation-interfacing fields of collaboration that can leverage physiology, engineering, biomechanics, robotics, and computer science, among others. We therefore propose a simple **Challenge-Mechanism-Partnership (CMP) framework** to help identify future collaborations.

The CMP framework begins with a research or conservation team identifying a conservation challenge (C) at the population, species, or ecosystem level. For example, a decline or shift in distribution is documented, a threat may be introduced or is increasing in severity, or some form of degradation has occurred. In response, the following question can be posed: "what mechanisms (M) can provide information on the cause and/or consequences of the conservation challenge"? For example, would having information on where organisms occur or how they move across a landscape; their stress, health, or nutritional status; how they grow or reproduce; behaviors under different contexts or in different locations; etc. improve the ability to understand the challenge or design a solution? In some cases, these mechanisms will fall within an existing conservation sub-field, such as those we have highlighted in our case studies. However, posing this question can also open up the opportunity to discover new subfields if the desired mechanistic information can be collected using tools from a field that is not yet conservation science-facing. Turning to the fields that have tools and techniques to measure or interpret these identified mechanisms will therefore require the formation of partnerships (P) with individuals from disciplines like computer science, engineering, robotics, physics, or mathematics. Given that all of the advances we encourage are dependent on collaboration, we also urge new partnerships be forged with ethical considerations, trust-building, and mutual respect at their center (e.g., Wilmer et al., 2021[66]), although a discussion of best practices in building interdisciplinary teams is beyond the scope of this manuscript. While we anticipate that the CMP approach has the biggest potential to identify new opportunities for integration in the conservation subfields we have detailed, as well as other long-standing mechanistic collaborations like conservation behavior, we are also optimistic that new subfields will emerge.

**Conclusion**

We have illustrated that there is considerable room for growth and integration of three mechanistic fields in conservation science: physiology, technology, and biomechanics. Our perspective has focused on biological disciplines and drawn on primarily Western science approaches, but we acknowledge that all of the proposed endeavors could be part of more holistic conservation programs. All of the fields we highlight require partnerships and provide the opportunity for working with individuals and/or organizations that may not have traditionally been conservation-facing. Overall, we advocate for individuals working on biological questions in conservation science to consider the value of mechanistic information, the potential it provides for expanding their monitoring and problem-solving approaches, and the capacity it can build for more interdisciplinary teams.

**Acknowledgements**
We acknowledge A. Patel for conversations regarding conservation biomechanics and cheetah rehabilitation. C. Madliger is supported by a *Natural Sciences and Engineering Research Council of Canada* Discovery Grant. A. Gouge was supported by a *Natural Sciences and Engineering Research Council of Canada* Undergraduate Student Research Award. We thank the open-source illustrators from Undraw.co, Bioicons, Phylopic, and NounProject for the images used in creation of the Figures.

**Data Availability**
The dataset compiled for this study is available from the corresponding author upon reasonable request.

**Author Contributions**
AKS and CLM conceptualized the manuscript. ASG performed the literature search and data compilation. AKS, ASG, and CLM wrote and edited the manuscript. AKS and ASG created the figures.